\documentclass[prd,twocolumn,fleqn,showpacs]{revtex4}
\usepackage[matrix,arrow]{xy}
\usepackage{graphicx}

\usepackage{amsmath}
\usepackage[psamsfonts]{amssymb}

\newcommand{\cH}{\ensuremath{\mathcal H}}

\newcommand{\cK}{\ensuremath{\mathcal K}}

\newcommand\PU{\operatorname{PU}}
\newcommand\Ind{\operatorname{Ind}}
\newcommand\Ca{$C^*$-algebra}
\newcommand\bbR{\mathbb R}

\newcommand{\CC}{{\mathbb C}}
\newcommand{\RR}{{\mathbb R}}
\newcommand{\ZZ}{{\mathbb Z}}
\newcommand{\bbZ}{{\mathbb Z}}
\newcommand{\bbC}{{\mathbb C}}

\newcommand{\TT}{{\mathbb T}}
\newcommand{\two}{\text{I}\text{I}}

\begin{document}
\def\thefootnote{\fnsymbol{footnote}}

\title{On mysteriously missing T-duals, H-flux and the T-duality group}

\author{Varghese Mathai$^1$ and Jonathan Rosenberg$^2$}
\affiliation{
$^1$Department of Pure Mathematics, University of Adelaide,
Adelaide, SA 5005, Australia \\
$^2$Department of Mathematics,
University of Maryland,
College Park, MD 20742, USA}

\begin{abstract}
\noindent
A general formula for the topology and H-flux of the T-duals  
of type {\two} string theories with H-flux on toroidal compactifications 
 is presented here.  
It is  known that
toroidal compactifications with H-flux do not necessarily have T-duals
which are 
themselves toroidal compactifications. A big puzzle has been to explain these
mysterious ``missing T-duals'', and our paper presents a solution 
to this problem using noncommutative topology. We also analyze 
the T-duality group and its action, and illustrate these concepts with examples.
\end{abstract}

%
\setcounter{footnote}{0}
\renewcommand{\thefootnote}{\arabic{footnote}}


\maketitle

\renewcommand{\thepage}{\arabic{page}}

T-duality is a symmetry of type {\two} string theories that
involves exchanging a theory compactified on a torus with a theory
compactified on the dual torus. 
The T-dual of a  type {\two} string theory compactified on a circle,
in the presence of a  
topologically nontrivial NS 3-form H-flux, was analyzed in special cases in
\cite{AABL,DLP,GLMW,KSTT}. There it was observed that T-duality
changes not only the H-flux, but also the spacetime topology.  
A general formalism for dealing with T-duality for
compactifications arising from a free circle action
was developed in \cite{bem}.  This formalism
was shown to be compatible with two physical constraints:
(1) it respects the local Buscher rules \cite{Bus},
and (2) it yields an isomorphism on twisted K-theory, in which
the Ramond-Ramond charges and fields take their values \cite{MM, Wib, BM}.
It was shown in \cite{bem} that T-duality exchanges the first Chern class 
with the fiberwise integral of the H-flux, thus giving a formula for 
the T-dual spacetime topology.
In this note we will present an account for physicists of the 
results in \cite{MR}, consisting of a  formula for  the T-dual
of a toroidal compactification, that is a theory compactified
via a free torus action, with  H-flux.  
One striking new feature that occurs for higher dimensional tori
is that not every toroidal compactification with H-flux has a T-dual;
moreover, even if  
it has a T-dual, then the T-dual need not be another 
toroidal compactification with H-flux.
A big puzzle has been to explain these
mysterious ``missing T-duals'', and our paper presents a solution 
to this problem using noncommutative topology. 
A similar phenomenon was noticed in \cite{LNR} in the 
special case of the trivial $\TT^2$ bundle over $\TT$
with non-trivial $H$-flux.
We also show that the generalized T-duality group $GO(n, n;\ZZ)$, $n$
being the rank 
of the torus, acts to generate the complete list of 
T-dual pairs related to a 
given toroidal compactification with H-flux. We will explain these
results by providing examples and applications.

In this letter we will consider type {\two}
string theories on target $d$-dimensional manifolds $X$, which  
are assumed to admit free, rank $n$ torus actions.
While for most physical applications one wants $d=10$, we do not need
to assume this, and in fact $X$ could 
represent a partial reduction of the original 10-dimensional spacetime
after preliminary compactification in $10-d$ dimensions.
The space of orbits of the torus action
on $X$ is given by a $(d-n)$-dimensional 
manifold, which we call $Z$.  
The freeness of the action implies that each orbit is  
a torus and that none of these tori degenerate.  As a result $X$ is
a principal torus bundle over the base $Z$, and so  
its topology is entirely determined by the topology of the base
$Z$ together with the first Chern class $c$ of the bundle 
$X \xrightarrow{p} Z$ in
$H^2(Z, \ZZ^n)$. This viewpoint is useful in that it automatically
identifies some gauge equivalent configurations, excludes
configurations not satisfying some equations of motion and imposes the
Dirac quantization conditions.
The Chern class $c$ is represented by a
vector valued closed 2-form with integral periods, the
curvature $F$. We will discuss conditions under which
the pair $(X \xrightarrow{p}Z,H)$ has a T-dual, either another pair
$(X \xrightarrow{p^\#}Z, {H}^\#)$ with the same base $Z$ (the
``classical'' case) 
or a more general non-commutative object (the ``nonclassical''
case). In both cases, there should be a sense in which string theory
on the original space $X$ (with H-flux $H$) is equivalent to a theory
on the T-dual.

\medskip

\noindent
\textbf{\textit{Basic setup:}
Let $p \colon X\to Z$ be a principal $T$-bundle
as above, where $T = (S^1)^n = \TT^n$ is a rank $n$ torus. Let
$H \in H^3(X, \ZZ)$ be an H-flux on $X$ satisfying
$\iota^* H = 0$, $\iota^*\colon H^3(X, \ZZ)\to H^3(T, \ZZ)$,
where $\iota \colon T \hookrightarrow X$ is the inclusion of a fiber. 
(This condition is automatically satisfied when $n\le 2$.)
}

\medskip
The  simplest case when the condition
$\iota^*(H)=0$ does {\em not} apply
is $X=\TT^3$, when considered as a rank 3, principal 
torus bundle over a point, with H-flux  a non-zero
integer multiple of the volume 3-form on $\TT^3$.
When $\iota^*(H)\ne 0$, there is no T-dual in the sense we are
considering, even in what we call the ``nonclassical'' sense.

It turns out that nontrivial bundles are always T-dual to 
trivial bundles with non-zero H-flux.  
Therefore we will need to include the fluxes $H$ and ${H}^\#$
in our toroidal compactifications, which are then
topologically determined by the triples $(Z,c,H)$ and
$(Z, c^\#, {H}^\#)$, where $H$ and
$H^\#$ are closed three-forms on the total spaces $X$ and $X^\#$
respectively.  

\medskip

\noindent
{\bfseries\textit{Our result on classical T-duals:}
Suppose that we are in the basic setup as above. 
Choose a basis $\{\TT^2_j\}_{j=1}^{k}, \; k=\binom n 2$ for
$H_2(T, \ZZ)$ consisting of 2-tori,  and push this forward into 
$H_2(X, \ZZ)$ via $\iota_*$.  We can consider the
cohomology classes 
\[
\int_{\TT^2_j} H = H \cap \iota_*(\TT^2_j) \in 
H^1(X, \ZZ).  
\]
These classes restrict to $0$ on the fibers, since
$\iota^*(H)=0$.  Using the following exact sequence, derived from the 
spectral sequence of the torus bundle,
\begin{equation}
0 \to H^1(Z, \ZZ) \xrightarrow{p^*} H^1(X, \ZZ)
\xrightarrow{\iota^*} H^1(T, \ZZ) \to \cdots,
\end{equation} 
we see that the classes $\int_{\TT^2_j} H = H \cap \iota_*(\TT^2_j)
\in H^1(X, \ZZ)$ 
come from unique classes $\{\beta_j\}_{j=1}^k$ in  $H^1(Z, \ZZ)$.
Set 
\begin{equation}
p_!(H) = \Bigl(\beta_1 , 
\, \ldots,  \beta_k \Bigr) \in H^1(Z, \ZZ^k).
\end{equation} 
If $p_!(H) = 0 \in H^1(Z, \ZZ^k)$, and in particular if $Z$ is simply
connected, then there is a classical T-dual to $(p, H)$, consisting of
$p^\# \colon X^\# \to Z$, which is another principal $T$-bundle
over $Z$, and $H^\# \in H^3(X^\#, \bbZ)$,
the T-dual H-flux on  $X^\#$. One obtains
a commuting diagram of the form
\begin{equation}
\label{eq:fibdiag}
\xymatrix{
 & X\times_Z X^\# \ar[ld]_{p^*(p^\#)} \ar[rd]^{(p^\#)^*(p)} & \\
X \ar[rd]_p& & X^\# \ar[ld]^{p^\#}\\
 & Z & .}
\end{equation}
In this case, the compactifications topologically  
specified by $(Z, c, H)$ and  
$(Z,c^\#,{H}^\#)$ are T-dual if 
$c, \,c^\# \in H^2(Z, \ZZ^n)$ are related as follows:

Let $c_j$, $j=1,\cdots,n$, be the components of $c$. 
Let $X_j  \xrightarrow{\pi_j} Z$ be the principal $\TT^{n-1}$ 
subbundle of $X$ obtained by deleting $c_j$, i.e.
the Chern class of $X_j$ is 
$$c(\pi_j) = 
(c_1, \ldots, \hat c_j, \ldots, c_n).$$
Then  $X \xrightarrow{p_j} X_j $ is a 
 principal $S^1$-bundle whose Chern class
 is equal to $\pi_j^*(c_j)$.
Define $X_j^\#  \xrightarrow{\pi_j^\#} Z$, $X^\# \xrightarrow{p_j^\#} X_j^\#$ 
etc. similarly.  Then we have
$$(\pi_j)^*(c_j^\#) = (p_j)_!(H)\quad {\rm \bf and}\quad 
(\pi_j^\#)^*(c_j) = (p_j^\#)_!(H^\#).$$
}

\medskip

Here the correspondence space $X\times_Z X^\#$ is the submanifold 
of $X\times X^\#$ consisting of pairs of points $(x, y)$ such that 
$p(x) = p^\#(y)$, and has the property that it 
implements the T-duality between $(p, H)$ and
$(p^\#, H^\#)$. It also turns out that $p^\#_!(H^\#)=0\in H^1(Z, \ZZ^k)$
and that the T-dual of $(p^\#, H^\#)$ is $(p, H)$. 
So in this case, T-duality \emph{exchanges the  integral of the H-flux }
(over a basis of circles in the fibers) \emph{with the first Chern class}.
The condition in the result above determines, 
at the level of cohomology, the curvatures $F$ and ${F}^\#$.  
However the NS field strengths are  
only determined up to the addition of a three-form on the base $Z$,  
because the integral of such a form over a basis of circles in the
fibers vanishes.  This settles a conjecture in \cite{bem}, and
was also considered by \cite{bhm}.

The simplest higher rank example is $X= S^2 \times \TT^2$, considered as 
the trivial $\TT^2$ bundle over $Z=S^2$, with H-flux 
equal to $H= k_1 a \wedge  b_1 +  k_2 a\wedge b_2$, where we use the K\"unneth
theorem to identify $H^3(S^2 \times \TT^2, \ZZ)$ with 
$H^2(S^2, \ZZ)\otimes H^1(\TT^2, \ZZ)$, and $a$ is the generator of 
$H^2(S^2, \ZZ)\cong \ZZ$, $b_1, b_2 $ are the generators of 
$H^1(\TT^2, \ZZ)\cong \ZZ^2$ and $k_1, k_2 \in \ZZ$. Since $S^2$ is simply connected,
$p_!(H) = 0$ and the T-dual of $(S^2 \times \TT^2, H)$ is the 
nontrivial rank 2 torus bundle $P$ over $S^2$ with Chern class 
$c_1(P) = (k_1 a, k_2 a) \in H^2(S^2, \ZZ) \oplus H^2(S^2, \ZZ)
= H^2(S^2, \ZZ^2)$, and with H-flux equal to zero.
This example generalizes easily by taking the Cartesian product 
with a manifold $M$, and pulling back the $H$-flux
to the product and arguing as before, we see that 
the T-dual of $(M \times S^2 \times \TT^2, H)$ is 
$(M \times P, 0)$.

\medskip

\noindent
\textbf{\textit{Our result on nonclassical T-duals:}
Suppose that we are in the basic setup as above. If
$p_! (H) \ne 0 \in H^1(Z, \ZZ^k),$ 
then there is {\em no} classical T-dual to
$(p, H)$; however, there is a nonclassical T-dual consisting of
a continuous field of (stabilized) noncommutative tori $A_f$ over $Z$,
where the fiber over the point $z\in Z$ is equal to the 
rank $n$ noncommutative torus $A_{f(z)}$
(see Figure 1 below).
Here $f \colon Z \to \TT^k$ is a continuous map representing $p_! (H)$.
}

\begin{figure}[ht]
\includegraphics[height=1.5in]{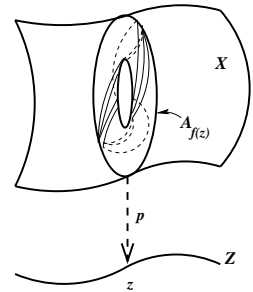}\\
\caption{In the diagram, the fiber over $z\in Z$ 
is the noncommutative torus $A_{f(z)}$,
which is represented by a foliated torus, with foliation 
angle equal to $f(z)$.}
\label{fig:folbundle}
\end{figure}

This suggests an unexpected
link between classical string theories and the ``noncommutative'' ones,
obtained by ``compactifying'' matrix theory on tori,
as in \cite{CDS} (cf.\ also \cite[\S\S6--7]{SW}). We now recall 
the definition of the 
rank $n$ noncommutative torus $A_\theta$, cf.\ 
\cite{Rieffel}. This algebra (stabilized by tensoring
with the compact operators $\cK$) occurs geometrically 
as the foliation algebra associated to Kronecker
foliations on the torus \cite{Connes}. 
In \cite{CDS}, the same algebra  occurs naturally 
from studying the field equations of the IKKT 
(Ishibashi-Kawai-Kitazawa-Tsuchiya) model compactified on
$n$-tori, or from the study of BPS states of the BFSS 
(Banks-Fisher-Shenker-Susskind) model.  (The IKKT and BFSS
models are both large-$N$ matrix models in which Poisson brackets in
the Lagrangian are replaced by matrix commutators.)
For each $\theta \in \TT^k$, identified with a hermitian matrix
$\theta = (\theta_{ij}),\; i, j =1,\ldots n$,
$\theta_{ij} \in S^1$ with $1$'s down the diagonal, 
the \emph{noncommutative torus} 
$A_\theta$ is defined abstractly 
as the $C^*$-algebra generated by $n$ unitaries
$U_j, \; j=1, \ldots , n$ in an infinite dimensional Hilbert space 
satisfying the commutation relation
$U_i U_j=  \theta_{ij} U_jU_i, \; i, j=1, \dots, n$. Elements in $A_\theta$ 
can be represented by infinite power series
\begin{equation}\label{schwartz}
f = \sum_{m\in \ZZ^n} a_{m} \, U^m,
\end{equation} 
where $a_{m} \in \bbC$ and $U^m = U_1^{m_1}\ldots U_n^{m_n}$, for all
$m = (m_1, \ldots, m_n)\in \ZZ^n$.

A famous example of a principal torus bundle with non-T-dualizable 
H-flux
is provided by $\TT^3$, considered as the trivial $\TT^2$-bundle over 
$\TT$,
with $H$ given by $k$ times the volume form on $\TT^3$.  $H$ is non 
T-dualizable
in the classical sense since $p_!(H)\ne 0 \in H^1(\TT, \ZZ)$. Alternatively,
there are no non-trivial principal $\TT^2$-bundles over $\TT$, since
$H^2(\TT,\ZZ^2) =0$, that is, there
is no way to dualize the H-flux
by a (principal) torus bundle over $\TT$, cf.\ \cite{KSTT}. This is an
example of a mysteriously missing T-dual.
This example is covered by our result on nonclassical T-duals above.
The T-dual is realized by a field of stabilized
{\em noncommutative tori} fibered over $\TT$.
Let $\cH = L^2(\TT)$ and consider the the projective unitary 
representation $\rho_\theta \colon \ZZ^2 \to \PU(\cH)$ 
in which the generator of the first $\ZZ$ factor 
acts by multiplication by $z^k$ (where $\TT$ is thought of as the unit circle 
in $\CC$) and the generator of the second $\ZZ$ factor acts by
translation by $\theta \in \TT$. 
Then the Mackey obstruction of $\rho_\theta$ is $\theta^k \in \TT \cong 
H^2(\ZZ^2, \TT)$.
Let $\cK(\cH)$ denote the algebra of compact operators on $\cH$
and define an action $\alpha$ of $\ZZ^2$ on continuous functions on the 
circle with values in compact operators, $C(\TT, \cK(\cH))$,  given at 
the point $\theta$ by $\rho_\theta$. Define the $C^*$-algebra $B$, which 
is obtained by inducing the $\ZZ^2$ action to an action of $\bbR^2$ on  
$B = {\Ind}_{\ZZ^2}^{\RR^2}\left(C(\TT, \cK(\cH)), \alpha\right)$, i.e. 
$ B = \{ f: \RR^2 \to C(\TT, \cK(\cH)):
  f(t+g) = \alpha(g)( f(t)),\;\; t\in \RR^2, g\in\ZZ^2 \}.$
Then $B$ is a continuous-trace {\Ca} having spectrum $\TT^3$
and Dixmier-Douady invariant $H$. $B$ also has
an action of $\RR^2$ whose induced action on the spectrum of $B$
is the trivial bundle $\TT^3\to \TT$.  Then our noncommutative T-dual 
is the crossed product algebra
$B\rtimes \RR^2 \cong C(\TT, \cK(\cH))\rtimes \ZZ^2 = A_f$ , which has fiber
over $\theta \in \TT$ given by
$\cK(\cH) \rtimes_{\rho_\theta} \ZZ^2 \cong 
A_\theta \otimes \cK(\cH,)$ where $A_\theta$ is the noncommutative $2$-torus.
In fact, the crossed product  $B\rtimes \RR^2$ is isomorphic
to the (stabilized) group $C^*$-algebra $C^*(H_{\bbZ} ) \otimes \cK$, where $H_{\bbZ}$ is the
integer Heisenberg-type group,
\begin{equation}
H_{\mathbb Z} = \left\{   \begin{pmatrix} 1 & x & \frac{1}{k}z \\
0 & 1 & y\\
0 & 0 &       1
\end{pmatrix} : x, y, z \in \mathbb Z \right\}.
\end{equation}
In summary, the nonclassical T-dual of $(\TT^3, H=k)$ is 
$A_f = C^*(H_{\bbZ} ) \otimes \cK$. As required in order to match up
RR charges, the $K$-theory of this algebra is the same as the
$K$-theory of $\TT^3$ with twist given by our H-flux,
or $k$ times the volume form.

This example generalizes easily by taking the Cartesian product 
with a manifold $M$. Pulling back the $H$-flux
to the product and arguing as before, we see that $(M \times \TT^3, H = k)$ 
is T-dual to $C(M)  \otimes C^*(H_{\bbZ} ) \otimes \cK$.
For instance, if the dimension of $M$ is seven, then  $ M \times
\TT^3$ is ten dimensional, yielding examples of spacetime manifolds
that are relevant to type {\two} string theory. 

\medskip

\noindent
{\textbf{\textit{Our results on the T-duality group:}}}

It is important to realize that a fixed space $X$ can 
sometimes be given the structure of a principal torus bundle over $Z$
in many different ways. For example, given a free action of a torus
$T=\TT^n$ on $X$, with quotient space $Z=X/T$, we can for every
element $g\in {\rm Aut}(\TT^n)=GL(n,\ZZ)$ define a 
new free action of $T$ on $X$, twisted by $g$, by the formula
$x\cdot_g t = x\cdot g(t)$. (Here $t\in T$, $\cdot$ is the
original free right action of $T$ on $X$, and
$\cdot_g$ is the new twisted action.) If $c\in H^2(Z,\ZZ^n)$ was the
Chern class of the original bundle, the Chern class of the $g$-twisted
bundle is $g\cdot c$, with $g$ acting via the action of $GL(n,\ZZ)$ on
$\ZZ^n$. 

The group $GL(n,\ZZ)$ embeds in $O(n,n;\ZZ)$, the subgroup of
$GL(2n, \ZZ)$ preserving the quadratic form defined by
$\begin{pmatrix}0&I_n\\I_n&0\end{pmatrix}$, via
$a\mapsto \begin{pmatrix}a&0\\0&(a^t)^{-1}\end{pmatrix}$ (see
\cite[\S2.4]{GPR}). This larger group $O(n,n;\ZZ)$
is often called the
\emph{T-duality group}. In fact we will consider the still larger
\emph{generalized T-duality
group} $GO(n,n;\ZZ) = O(n,n;\ZZ)\rtimes (\ZZ/2)$ of matrices in $GL(2n,
\ZZ)$ preserving the form $\begin{pmatrix}0&I_n\\I_n&0\end{pmatrix}$ up
to sign. Good references for the T-duality group
include \cite{GPR} (for the state of the theory up to 1994)
and \cite{Hull} for more current developments. 

{\bfseries Suppose that we are in the basic setup as above, with $Z$ simply
connected, so that one is always guaranteed to have a classical
T-dual. Then the generalized T-duality group $GO(n, n;\ZZ)$ acts on 
the set of T-dual pairs $(p, H)$ and $(p^\#, H^\#)$ to generate all
related T-dual pairs.  \emph{All of these pairs are physically
equivalent.} The restriction of the action to $GL(n,\ZZ)$ (as embedded
above) corresponds to twisting of the action 
on the same underlying space as above. 

When $Z$ is not simply connected and
$p_!(H)\ne 0$, it is not clear that one has an action of
the full T-duality group. But the action of $GL(n,\ZZ)$  always sends
the pair consisting of $(p, H)$ and its nonclassical T-dual to another
nonclassical pair, involving continuous fields of (stabilized)
noncommutative tori over $Z$.
}

\medskip

We illustrate the action of the {generalized T-duality group} 
in the simplest case of circle bundles with H-flux, 
in which case the generalized T-duality group reduces to 
$GO(1, 1;\bbZ)$, a dihedral group of oder $8$.

Consider the example of the 3 dimensional 
lens space $L(1,p)=S^{3}/\ZZ_p$, with $H$-flux 
$H= q$ times the volume form, cf.\ \cite{MMS}. 
Here $p, q \in \bbZ$, and initially we take $p,\,q> 0$. Then
$L(1,p)$  is a circle bundle 
over the 2-dimensional sphere $S^2$ and has first Chern class
equal to $p$  times the volume form of $S^2$. Then, as shown in
\cite[\S 4.3]{bem}, $(L(1,p),H=q)$ and $(L(1,q),H=p)$ are T-dual to 
each other, and the element $\begin{pmatrix}0 & 1\\ 1& 0\end{pmatrix}$
of $O(1,1; \bbZ)$ interchanges them. The element $\begin{pmatrix}-1 &
0\\ 0& -1\end{pmatrix}$ of the T-duality group 
$O(1,1; \bbZ)$ lies in the subgroup $GL(1,\ZZ)$, embedded as above,
and acts by twisting the $S^1$ action on $L(1, p)$.
This twisted action makes $L(1, p)$ into a circle bundle
over $S^2$ having first Chern class
equal to $-p$ times the volume form of $S^2$. This bundle is denoted
$L(1, -p)$, and its total space is diffeomorphic to $L(1, p)$,
though by an orientation-reversing diffeomorphism.  Therefore 
the action of $\begin{pmatrix}-1 & 0\\ 0& -1\end{pmatrix}$ on the pair
$(L(1,p),H=q)$ and $(L(1,q),H=p)$ gives rise to a new T-dual pair
$(L(1,-p),H=-q)$ and $(L(1,-q),H=-p)$. The group $GO(1,1; \bbZ)$ is
generated by the two elements of $O(1,1; \bbZ)$ just discussed and by
$\begin{pmatrix}1 & 0\\ 0& -1\end{pmatrix}$, which replaces the
original T-dual pair by the pair consisting of
$(L(1,p),H=-q)$ and $(L(1,-q),H=p)$. Here we have tacitly assumed
$p,\,q\ge 2$; we can extend things to other values of $p$ and $q$
by making the convention that $L(1,1) = S^3$ and $L(1,0)
= S^2 \times S^1$. This refines the T-duality in \cite[\S 4.3]{bem}.
Thus in general there are 8 different (bundle, H-flux) pairs with
equivalent physics, corresponding to $(\pm p, \pm q)$ and
$(\pm q, \pm p)$.

This example generalizes easily by taking the Cartesian product 
with a manifold $M$.
For instance, if the dimension of $M$ is seven, then we obtain
8 different (bundle, H-flux) pairs in the same  $GO(1,1; \bbZ)$-orbit
as $ M \times L(1, p)$. All of these are  ten-dimensional spacetime
manifolds relevant to type {\two} string theory.

We end with some open problems.
A critical verification of any proposed duality is that the anomalies 
should match on  
both sides.  This was checked for T-duality involving circle bundles 
with H-flux in 
\cite{bem}, but remains to be analyzed in the general torus bundle case
with H-flux. It also remains to be determined whether or not the 
group $GO(n, n;\ZZ)$ also operates in the nonclassical case.
Another problem is to extend our results to non-free torus
actions \cite{DP}, in which case it could be relevant to mirror symmetry.

\noindent
{\bf Acknowledgments:}

\noindent
VM was financially supported by the Australian Research Council
and JR was partially supported by NSF Grant DMS-0103647.

\end{document}